\nonstopmode
\documentclass[aip,pop,reprint,superscriptaddress]{revtex4-2}

\usepackage{float}
\usepackage{graphicx}
\usepackage{epstopdf}
\usepackage{amsmath,amssymb,amsfonts}
\usepackage[usenames,dvipsnames]{color}
\usepackage{subcaption}
\usepackage[normalem]{ulem}

\begin{document}

\title{Laboratory evidence of electron pressure anisotropy driving %the growth rate of 
plasmoid mediated magnetic reconnection %in %laser-produced 
%plasmas %with high-aspect ratio current sheet 
}

\author{A. Sladkov}
\affiliation{Light Stream Labs LLC, USA, Palo Alto, CA 94306}
\author{T. Waltenspiel}
\affiliation{LULI - CNRS, CEA, UPMC Univ Paris 06 : Sorbonne Universit\'e, Ecole Polytechnique, Institut Polytechnique de Paris - F-91128 Palaiseau cedex, France}
\affiliation{INRS-EMT, 1650 boul, Lionel-Boulet, Varennes, QC, J3X 1S2, Canada}
\affiliation{University of Bordeaux, Centre Lasers Intenses et Applications, CNRS, CEA, UMR 5107, F-33405 Talence, France}
\author{H. Ahmed}
\affiliation{Central Laser Facility, Rutherford Appleton Laboratory,  Didcot OX11 0QX, UK4}
\author{A. Alexandrova}
\affiliation{LULI - CNRS, CEA, UPMC Univ Paris 06 : Sorbonne Universit\'e, Ecole Polytechnique, Institut Polytechnique de Paris - F-91128 Palaiseau cedex, France}

\author{V. Anthonippillai }
\affiliation{LULI - CNRS, CEA, UPMC Univ Paris 06 : Sorbonne Universit\'e, Ecole Polytechnique, Institut Polytechnique de Paris - F-91128 Palaiseau cedex, France}
\author{P. Antici}
\affiliation{INRS-EMT, 1650 boul, Lionel-Boulet, Varennes, QC, J3X 1S2, Canada}
\author{S. N. Chen}
\affiliation{Light Stream Labs LLC, USA, Palo Alto, CA 94306}
%\affiliation{ELI-NP, ``Horia Hulubei'' National Institute of Physics and Nuclear Engineering, Bucharest - Magurele, Romania}
\author{I. Cohen}
\affiliation{Technion Israel Institute of Technology, Faculty of Physics, Technion City, Haifa 3200003,
Israel}
\author{E. d'Humières} 
\affiliation{CELIA, Universit\'e de Bordeaux–CNRS–CEA, UMR 5107, 33405 Talence, France}

\author{W. Yao}
\affiliation{LULI - CNRS, CEA, UPMC Univ Paris 06 : Sorbonne Universit\'e, Ecole Polytechnique, Institut Polytechnique de Paris - F-91128 Palaiseau cedex, France}
\affiliation{Sorbonne Universit\'e, Observatoire de Paris, Universit\'e PSL, CNRS, LUX, F-75005, Paris, France}
\author{J. Fuchs}
\email{Author to whom correspondence should be addressed:  julien.fuchs@polytechnique.edu, julien.fuchs@technion.ac.il}
\affiliation{LULI - CNRS, CEA, UPMC Univ Paris 06 : Sorbonne Universit\'e, Ecole Polytechnique, Institut Polytechnique de Paris - F-91128 Palaiseau cedex, France}
\affiliation{Technion Israel Institute of Technology, Faculty of Physics, Technion City, Haifa 3200003,
Israel}

\begin{abstract}

Plasmoid-driven magnetic reconnection in elongated current sheets is suspected to be an ubiquitous phenomenon in space and astrophysical plasmas, but the mechanisms driving its onset and dynamics are still debated. Deciphering the  physical  mechanisms  dominating  the  destabilization  and  fragmentation of  the  current sheet, as well as its evolution, would have a wide impact into our understanding of the induced plasma turbulence and particle acceleration. Here, by coupling 3D hybrid simulations with laser-driven experiments that involve counterflowing high-energy-density magnetized plasmas with a long aspect ratio of their contact layer, we show that electron pressure anisotropy is the driving factor of the growth rate of the tearing instability, and will sustain the reconnection process even without classical resistivity. Dissipative mechanisms, such as resistivity and isotropization, are further found to stabilize the sheet to varying degrees, thus modifying plasmoid formation. By identifying the roles of pressure anisotropy, dissipation, and large-scale geometry, our work  lays the groundwork for the evaluation of plasmoid-driven reconnection impact on the dynamics of laboratory and astrophysical plasmas.

%induced in these events.  (classical  resistivity,  Hall  effects,  pressure anisotropy, or other kinetic processes), and how these in-gredients quantitatively determine the reconnection rate,plasmoid growth, and energy partition.

%We investigate magnetic reconnection dynamics in counterflowing high-energy-density magnetized plasmas having a long aspect ratio of their contact layer. %The experiments focus on anisotropic plasma plumes with large aspect ratios, providing a novel platform for exploring reconnection in systems with extended current sheets.
 %Complementary hybrid simulations, which incorporate ion dynamics and electron pressure tensor effects, reproduce the key features of reconnection in this regime. %Synthetic proton radiographs generated from the model show excellent agreement with experimental observations, confirming plasmoid formation. 
%The simulations  suggest that electron pressure anisotropy drives fast tearing instabilities, sustaining reconnection even without classical resistivity. Dissipative mechanisms (resistivity and isotropization) stabilize the sheet to varying degrees, modifying plasmoid formation. Our findings extend earlier work by demonstrating plasmoid-mediated reconnection in elongated current sheets, thereby offering insights relevant to astrophysical systems and to long-duration laboratory experiments.
\end{abstract}

\maketitle

\section{Introduction} \label{sec:intro}

% Magnetic reconnection~\cite{taylor1986} is the process that occurs when two plasmas having anti-parallel magnetic field-lines 
% meet each other. As magnetic field is annihilated, the magnetic field energy can be converted into the energy of charged plasma particles. 
Magnetic reconnection~\cite{taylor1986} is the process that occurs in nature when two plasmas carrying oppositely directed magnetic fields interact. During this process, the magnetic field is reorganized, and part of its energy is converted into the kinetic and thermal energy of the plasma particles. Magnetic reconnection can take place under very different conditions and scales, and is evidenced from the laboratory~\cite{taylor1986} to astrophysics~\cite{degouveia2010}. %It is particularly interesting when it comes to astrophysics as plasma constitutes 99 \% of the visible matter in the Universe. 
In the latter case, magnetic reconnection %can 
occurs in a wide range of configurations, from stellar eruptions creating  arches and coronal mass ejection~\cite{Xue2020}, to %interaction between 
stellar winds interacting with planetary magnetic field~\cite{Graham2022}, thereby shaping the planet's magnetosphere and magnetotail~\cite{Nishida1990}, where magnetic field lines reconnect. In this aforementioned list, %the case 
the most common type of reconnection configuration %most likely to happen 
consists in the encountering plasmas interacting %in 
over an elongated %and large ratio 
zone of space, thereby creating a current sheet. This current sheet becomes subject to magnetic stress, when the frozen-in conditions in plasmas \cite{Syrovatskii1971} are disrupted. Then, where the magnetic field lines meet, this induces the formation of a chain of multiple reconnection sites~\cite{Alexandrova2015},  leading to the formation of plasmoids ~\cite{Daldorff2022}.

Plasmoid-driven reconnection can be measured in space by measuring, for example, the accelerated particles or plasma emission from plasmoids %directly from the sheet
~\cite{Palmroth2023}. However, even though plasmoids are understood to exist, it is still difficult to make measurements meaningful \textit{in situ}~\cite{Li2016}, as they remain spontaneous, succint, hardly previsible, and therefore rarely detected. 

Hence, it is presently extremely difficult to infer from these measurements the physics governing the onset and evolution of plasmoid-mediated reconnection in extended current sheets, namely what physical mechanisms dominate the destabilization and fragmentation of the sheet (classical resistivity, Hall effects, pressure anisotropy, or other kinetic processes), 
and how these ingredients quantitatively determine the reconnection rate, plasmoid growth, and energy partition.

Here, we present the results of an experiment performed at the LULI2000 laser facility (France), aimed at investigating these onset and evolution of plasmoid mediated reconnection. The interest of laboratory experiments is that they not only yield global observations of the magnetic process with high spatial resolution, but also allow to access its temporal dynamics, in order to help decipher the underlying microphysics. %  Our work is then part of the study of magnetic reconnection in laboratory in order to reproduce space plasma environments and detail the process.
We used two counter-propagating and anti-parallel elongated magnetic field toroids, from which we induced magnetic reconnection with a long aspect ratio current sheet. Time-resolved proton radiography~\cite{Schaeffer2023} captures the formation, instability, and fragmentation of the current sheet into plasmoids. Experimentally, a first phase of magnetic field compression is observed to take place, followed by the fragmentation of the current sheet in dynamically evolving plasmoids, to end up in a saturation phase where filaments elongate and stabilize. 
Complementary 3D hybrid simulations performed with the AKA code (see Appendix A - Section~\ref{sec:model}), which incorporate ion dynamics and electron pressure tensor effects, confirm that elongated current sheets fragment into plasmoids - in agreement with experimental proton radiographs. Our main finding, derived from matching the experimental observations to the simulations, is that electron pressure anisotropy drives fast tearing instabilities, sustaining reconnection even without classical resistivity. This acts in competition with dissipative mechanisms (resistivity and isotropization), which stabilize the sheet to varying degrees, modifying plasmoid formation.

The present experiment builds on the progress made in the last few decades, using laser-produced plasmas, where magnetic fields can be self-generated via the Biermann battery effect~\cite{biermann1951}. This has provided new opportunities for studying the reconnection process~\cite{nilson2006} in systems with a large ratio between the global size and the kinetic scales of the plasmas. As of yet, most of the experimental works on High Energy Density (HED) plasmas have focused on strongly forced reconnection in a single X-line~\cite{nilson2006,li2007,fox2011,fox2012,rosenberg2015,palmer2019, Bolanos2022}, with some experiments hinting at plasmoid formation~\cite{dong2012}. However, naturally found magnetic reconnection events typically develop from elongated current sheets, in which reconnection arises at multiple sites separated by plasmoids~\cite{Uzdensky2010}. Recently, several groups reported  first investigations of reconnection in HED plasmas having highly extended current sheets, the length of which far exceeded the internal kinetic scales of the plasma, such as particle inertial lengths and their gyroradii~\cite{Hare2017,fox2020, fox2021, pearcy2024}. 

Most recently, Pearcy et al.~\cite{pearcy2024} reported the first direct experimental observation of plasmoids in high-$\beta$ (i.e., the ratio of electron pressure to magnetic field pressure $\sim 10$) magnetic reconnection using laser-driven plasmas from %($\sim1$~ns) 
\text{CH} targets, albeit in a X-line configuration of two hemispherical colliding plasmas. Using high-resolution proton radiography and Thomson scattering, they identify multiple magnetic islands forming in a dynamically evolving current sheet with super-Alfvénic inflows. The current sheet thins below the ion skin depth, indicating reconnection in the Hall MHD regime~\cite{birn2001}. Their results confirmed that plasmoid-mediated reconnection enhances reconnection rates even in high-$\beta$ environments, consistently with modern theoretical predictions~\cite{Uzdensky2010}.

In numerical work, current sheets with aspect ratios above a critical value ($A=L/2\delta$ $\sim 50$, where $L$ is the current sheet length and $\delta$ is the sheet half-width), have been shown to be unstable to tearing modes and generate secondary islands ~\cite{loureiro2005}. Fast magnetic reconnection models based on the resistive tearing instability~\cite{loureiro2007} predict the formation of plasmoids whose number and properties depend on the Lundquist number $S = V_AL/\eta$, where $V_A$ is the upstream Alfv\'en speed and $\eta$ is the plasma resistivity. In  recent experimental works~\cite{fox2020, fox2021} magnetic reconnection was investigated in extended current sheets, in which $A$ exceeded $10^2$,  and the Lundquist number exceeded $10^3$. It  was observed~\cite{fox2021} that the classical (resistive)~\cite{furth1963} and the collisionless tearing~\cite{coppi1966} theories could not explain the observed plasmoid growth rates, as well as the fastest-growing mode properties. It was also suggested that the observed high reconnection rate is driven by electron momentum transport, which is itself regulated by the electron pressure tensor term in Ohm's law~\cite{fox2020}. The ion anisotropic tearing, which could also be responsible for the observed instability of the current sheets~\cite{fox2021},  may also  contribute to space physics phenomena such as magnetospheric substorm onset.

The paper is organized as follows: in Section~\ref{sec:exp}, we present the setup of the experiment and the experimental characterization of the observed magnetic reconnection event developing in a long aspect-ratio current sheet. In Section~\ref{sec:radio} we present the  synthetic observations, obtained from 3D simulations, of the reconnection event that correspond to the experimental ones. In Section~\ref{sec:diss}, we analyze in detail the results of the simulation, and the importance of the electron pressure anisotropy in driving the dynamics of the observed plasmoid-mediated reconnection event. A summary and conclusion are presented in Section~\ref{sec:conc}.

\begin{figure}
\includegraphics[width=0.48\textwidth]{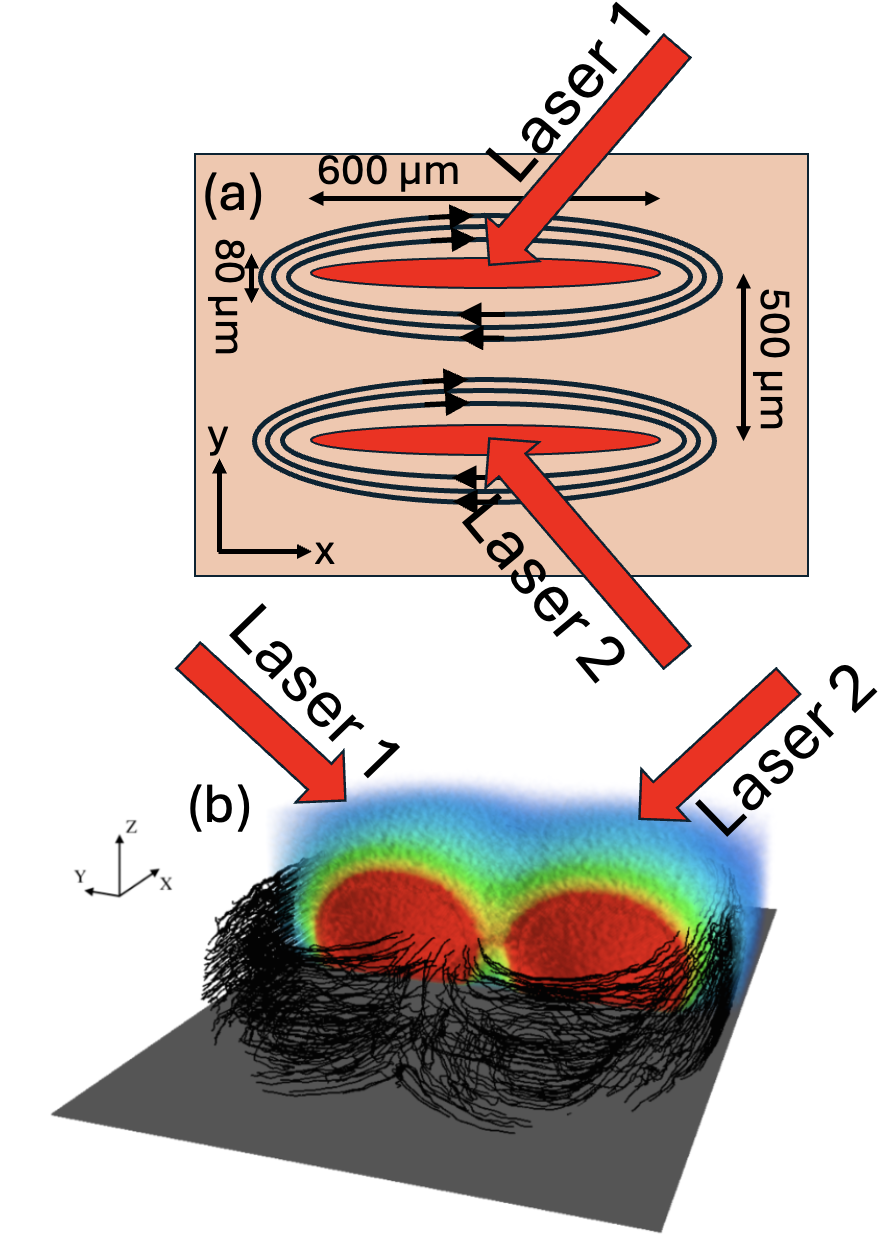}
\caption{(Color online) (a) Overall geometry of the experiment and (b) setup of the simulations. Two plasmas are generated following the ablation, using two high-power lasers, of a solid target. In each plasma,  magnetic fields are self-generated from the Biermann battery effect, as indicated by the black lines in panels (a-b). The encounter of the two plasmas, following their radial expansion along the target surface (in the x-y plane), will lead to the encounter of the anti-parallel magnetic fields,  to their reconnection, and to their eventual fragmentation into plasmoids. The image in panel (b) shows a snapshot from the numerical model with color-coded plasma density.%; overlaid black lines represent self-generated magnetic field lines arising from the Biermann battery effect. %Two laser beams irradiate copper targets at a separation of $500~\mu$m, producing counter-propagating anisotropic plumes. The long side of the elliptic focal spot ($\sim 600~\mu$m) and short ($\sim 80~\mu$m) of each plume establish a high aspect-ratio geometry that forces the formation of a narrow, elongated current sheet in the mid-plane. This initial configuration provides the conditions under which magnetic reconnection develops and eventually fragments into plasmoids.
} \label{fig:init}
\end{figure}

\section{Experiment} \label{sec:exp}

In our experiment, we study magnetic reconnection from laser-produced plasmas generated from copper (\text{Cu}) targets irradiated by 5~ns duration (square) laser pulses, each having an intensity of $10^{14}~\text{W/cm}^2$. A key distinguishing feature of our setup, as illustrated in Fig.~\ref{fig:init}.a, is the use of highly anisotropic focal spots, having dimension of approximately $600~\mu\text{m}$ by  $80~\mu\text{m}$, for the laser to deposit their energy onto the flat target. The laser focal spots are parallel to each other along the long dimension (the x-axis), and are separated by $500~\mu\text{m}$ along the y-axis. This irradiation geometry leads to elongated (along the x-axis) plasma plumes. The plasmas, being hot, the two plumes expand rapidly toward each other. As a result, the self-generated magnetic fields in each plume encounter in an anti-parallel geometry, leading to reconnection~\cite{nilson2006} over a long site, as illustrated in Fig.~\ref{fig:init}.a 

\begin{table}[h!]
\centering
\caption{Key physical parameters of the plasma within the current sheet region. Parameters above the line are measured or estimated; those below are derived.}
\vspace{0.5em}
\begin{tabular}{lc}
\hline
\textbf{Parameter} & \textbf{Approximate Value} \\
\hline
Plasma density, $n_e$ & $\sim 10^{19}~\text{cm}^{-3}$ \\
Average charge number, $Z$ & $\sim 19$ \\
Electron temperature, $T_e$ & $\sim 100~\text{eV}$ \\
Laser pulse duration & $5~\text{ns}$ \\
Laser intensity & $10^{14}~\text{W/cm}^2$ \\
Laser focal spot along the x-axis  & $600~\mu\text{m}$ \\
Laser focal spot along the y-axis  & $80~\mu\text{m}$ \\
Focal spot separation & $500~\mu\text{m}$ \\
\hline
Magnetic field, $B$ & $\sim 8{-}15~\text{T}$ \\
Plasma beta, $\beta$ & $\sim 2{-}6$ \\
Ion skin depth, $d_i$ & $\sim 60{-}120~\mu\text{m}$ \\
Electron skin depth, $d_e$ & $\sim 2{-}4~\mu\text{m}$ \\
Sound speed, $c_s$ & $\sim 5{-}10 \times 10^{4}~\text{m/s}$ \\
Alfvén speed, $v_A$ & $\sim 2{-}5 \times 10^{4}~\text{m/s}$ \\
Lundquist number, $S$ & $\sim 1.5{-}10^2$ \\
Current sheet width, $\delta$ & $\sim 100{-}200~\mu\text{m}$ \\
Aspect ratio, $L/\delta$ & $\sim 8{-}10$ \\
\hline
\end{tabular}
\label{tab:experiment_parameters}
\end{table}

The irradiation of copper targets, at the moderate intensity we use, generate plasmas, within the irradiated area, with electron densities on the order of $n_e \sim 10^{20}~\text{cm}^{-3}$ and temperatures in the range $T_e \sim 300$~eV, as inferred from Thomson scattering measurements (see Appendix C). Now, within the region of the current sheet, the plasmas are less hot and dense, as detailed in Table~\ref{tab:experiment_parameters} (see also Appendix C). Based on these parameters, we estimate the plasma to be near or slightly above $\beta \sim 5$, with magnetic fields in the range of $8{-}16$~T, ion skin depths $d_i \sim 60{-}120~\mu$m, and Lundquist numbers $S \sim 1{-}10^2$. The current sheet width, inferred from the measurements (see below), lies in the range $\delta \sim 100{-}200~\mu$m, corresponding to aspect ratios $L/\delta \sim 5{-}10$.

\begin{figure}
\includegraphics[width=0.48\textwidth]{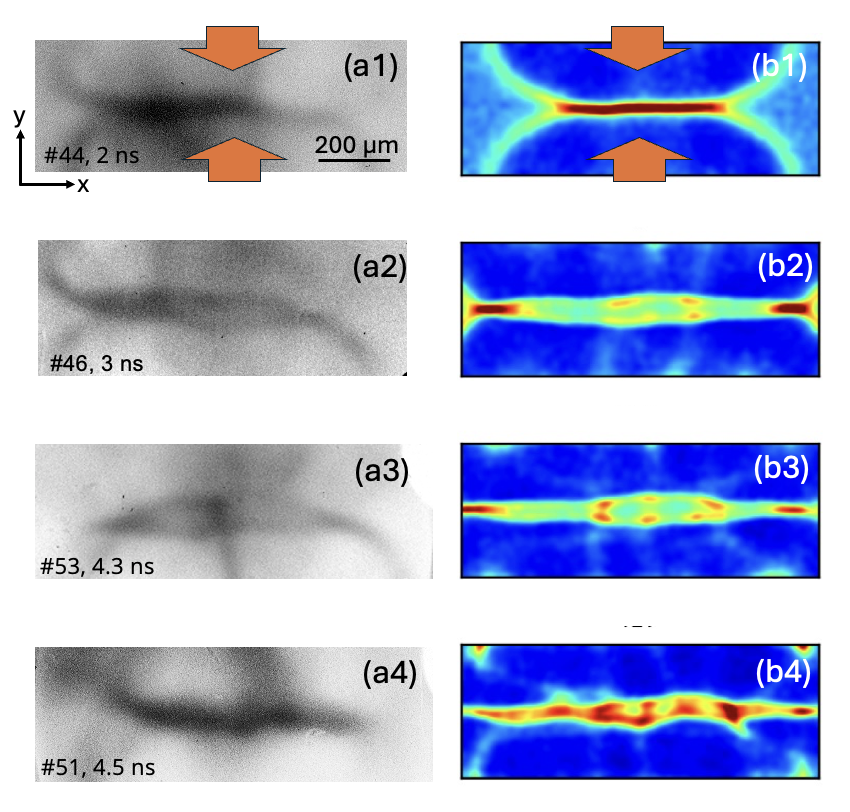}
\caption{ Time-resolved (a1-a4) experimental and (b1-b4) synthetic proton radiography of the reconnection process taking place with the current sheet (a) at $t=2$ ns, (b) at $t=3$ ns, (c) at $t=4.3$ ns, and (d) at $t=4.5$ ns. The orange arrows in panels a1 and b1 indicate the direction of the two plasma inflows in the reconnection region. The horizontal scale in panel (a1) applies to all images and is relative to the plane of the target in which the plasmas are produced. }\label{fig:proto_exp}
\end{figure}

The topology of the magnetic fields produced during and following the encounter of the two plasmas is analyzed using proton radiography~\cite{Schaeffer2023}. This technique uses fast (MeV) protons probing the plasma along the z-axis. After being deflected by the Lorentz force associated with the magnetic fields within the xy-plane,  they are collected on films \cite{bolton2014instrumentation}. The modulations of the proton dose on the film record the line-integrated deflections incurred on the protons as they propagated through the plasmas. The main information regarding the interaction between the two plasma plumes is contained in the pattern at the contact region, since this pattern is the result of the path-integrated magnetic field. In the cases with plasmoids, the modulations affecting the magnetic field in the contact region should lead to modulations on the trajectories of the probing protons. 

Figure~\ref{fig:proto_exp}.a1-a4 present a time-resolved sequence of proton radiographs capturing the dynamic evolution of the magnetic reconnection process as it occurred. The progression can be interpreted in terms of three distinct phases. In the \textit{initial phase} (a1), the two plasma flows converge and form a narrow, monolithic current sheet, seen as a well-defined, dark, elongated structure; this corresponds to the onset of magnetic reconnection, when the two anti-parallel fields are compressed against each other. In the \textit{fragmentation phase} (a2-a3), the current sheet becomes unstable to tearing, resulting in the formation of multiple localised features—interpreted as plasmoids—evident due to the breakup of the central pattern. Finally, the system enters a \textit{saturation phase} (a4), where plasmoid structures stabilise in size and shape, indicating a transition to nonlinear reconnection dynamics. This sequence provides direct visual evidence of current sheet formation, instability, and plasmoid saturation in a high-aspect-ratio geometry. The transition from a  thin and stable current sheet, at early times, to one that is fragmented into  plasmoids  is also witnessed by measuring the emissivity of the plasma in the optical visible domain, as detailed in  Appendix D - Section~\ref{sec:pyrometry}.

\section{Analysis in light of synthetic proton radiographs}\label{sec:radio}

To complement the experimental observation of magnetic reconnection and plasmoid formation, we perform AKA hybrid simulations that resolve ion kinetics while treating electrons as a fluid with a ten-moment pressure tensor closure (see Appendix A - Section~\ref{sec:model}). This approach captures key non-ideal effects, such as electron pressure anisotropy and Hall dynamics, which are essential for initiating and sustaining tearing instabilities in weakly collisional plasmas. By comparing early, intermediate, and late stages of the current sheet evolution, we can examine how small-scale instabilities develop, propagate, and influence the global magnetic topology, providing a detailed picture of plasmoid-mediated reconnection that complements and interprets the integrated experimental images of Figure~\ref{fig:proto_exp}.

For visualization of the temporal evolution of the magnetic fields and direct comparison with the experimental data we perform synthetic proton radiography. The synthetic proton radiographs are generated by a particle-tracing module of the  AKA code, using the field maps generated by the simulation and propagating the protons within. We model the fast laminar monoenergetic proton beam used in the experiment as originating from a point-like source, following which it crosses the interaction region. %Passing by protons are deflected by the electromagnetic field of the magnetized plasmas, after which they are collected on radiochromic films (RCF). 
The synthetic proton radiographs obtained from the simulation performed at Lundquist number $S \sim 100$ are shown in Figure~\ref{fig:proto_exp}.b1-b4. %The corresponding line-integrated  magnetic field maps from the simulations are shown in Figure  ~\ref{fig:bpath_proto}.%(b1-b4) display the proton fluence (dose) modulations impaired on protons for various configurations. 

%\begin{figure}
%\includegraphics[width=0.48\textwidth]{figs/proto_model.png}
%\caption{%Synthetic proton radiography from hybrid modeling at Lundquist number $S \sim 100$. Left column (a1–a4): 
%Path-integrated magnetic field distributions at four times ($t\Omega_0 = 15, 22.5, 30,$ and $37.5$), corresponding to the snapshot proton radiographs shown in Figure~\ref{fig:proto_exp}.% Right column (b1–b4): corresponding synthetic proton radiographs, displaying fluence modulations imposed on the probing beam. 
%\textcolor{red}{remove the synthetic p-rad, since we show them already in Fig2}}\label{fig:bpath_proto}
%\end{figure}

%Figure~\ref{fig:bpath_proto} presents the temporal evolution of synthetic proton radiographs derived from the hybrid simulations at Lundquist number $S \sim 100$. 
These radiographs provide a direct link between the simulated electromagnetic fields and the experimentally observable proton fluence patterns. The evolution proceeds as follows: 

\begin{enumerate}
    \item At $t\Omega_0 = 15$ (1-1.5ns), the central pattern appears as a thin line with only minor perturbations inside, indicating that the current sheet remains largely continuous and stable.
    \item By $t\Omega_0 = 22.5$ (1.5-2.5ns), the central line widens and transforms into a mouth-shaped structure. Small-scale perturbations emerge within the widened region, signaling the initial stages of current sheet destabilization and the onset of plasmoid formation.
    \item At $t\Omega_0 = 30$ (2.5-3.5ns), %the central pattern begins to narrow, while small filament-like structures sprout perpendicular to the main line. This reflects 
    the nonlinear development of tearing instabilities induces %and 
    the fragmentation of the sheet into multiple magnetic islands.
    \item Finally, at $t\Omega_0 = 37.5$ (3.5-4.5ns), the central line becomes more pronounced with visible intensity maxima, and the filaments grow longer and more prominent, illustrating the fully developed plasmoid-mediated reconnection stage.
\end{enumerate}

Overall, these results  clearly demonstrate the progressive destabilization of the current sheet and the emergence of plasmoids over time. Early-time smooth patterns confirm the initial quasi-stability of the sheet, while later-time widening, mouth-shaped structures, and filament-like sprouts reveal the dynamics of tearing instabilities and plasmoid evolution. However, because the radiographs are integrated along the proton path, the apparent number and size of plasmoids do not directly correspond to the actual 3D structures, as plasmoids may coalesce or break up at different heights (i.e. along the z-axis). Despite this limitation, the radiographs provide robust qualitative information about the development and evolution of magnetic reconnection, and the close agreement between the experimental and synthetic observations validates the hybrid modeling approach as a reliable tool for interpreting plasmoid-mediated dynamics in high-energy-density plasmas.

\section{Discussion}\label{sec:diss}

In classical resistive-MHD theory, tearing instabilities are enabled by non-ideal dissipation mechanisms that break the frozen-in condition of the magnetic flux~\cite{furth1963}. While the free energy driving the reconnection originates from the release of the magnetic tension stored in the current sheet, resistivity (or other generalized forms of dissipation) provides the mechanism that allows field lines to change topology and magnetic energy to be converted into flows and particle heating. In collisional plasmas, Ohmic diffusion plays this role, leading to the well-known scaling of the tearing growth rate with the Lundquist number $S$~\cite{furth1963, biskamp1986}. In weakly collisional or collisionless plasmas, such as here, alternative non-ideal effects such as the Hall term, electron inertia, and off-diagonal components of the electron pressure tensor replace classical resistivity as the enabling agent for reconnection~\cite{shay1999, birn2001, daughton2009}.

In the collissionless case, a key aspect of the reconnection dynamics is that the only available source of free energy in the electron diffusion region is the buildup of the electron pressure anisotropy leading to the development of a Weibel-type instability. The linear analysis of our hybrid model  presented in Appendix B (Section~\ref{sec:app}) shows that the growth rate of the anisotropy-driven modes scales as $\gamma \propto \sqrt{1-A}$, where $A = P_{xx}/P_{zz}$ ($P_{zz} > P_{xx}$) quantifies the anisotropy level, and that it stays on electron time scales. Thus, the development and persistence of electron pressure anisotropy is central to both the fast initiation of tearing and the subsequent nonlinear evolution of reconnection in this regime.

\begin{figure}
\includegraphics[width=0.48\textwidth]{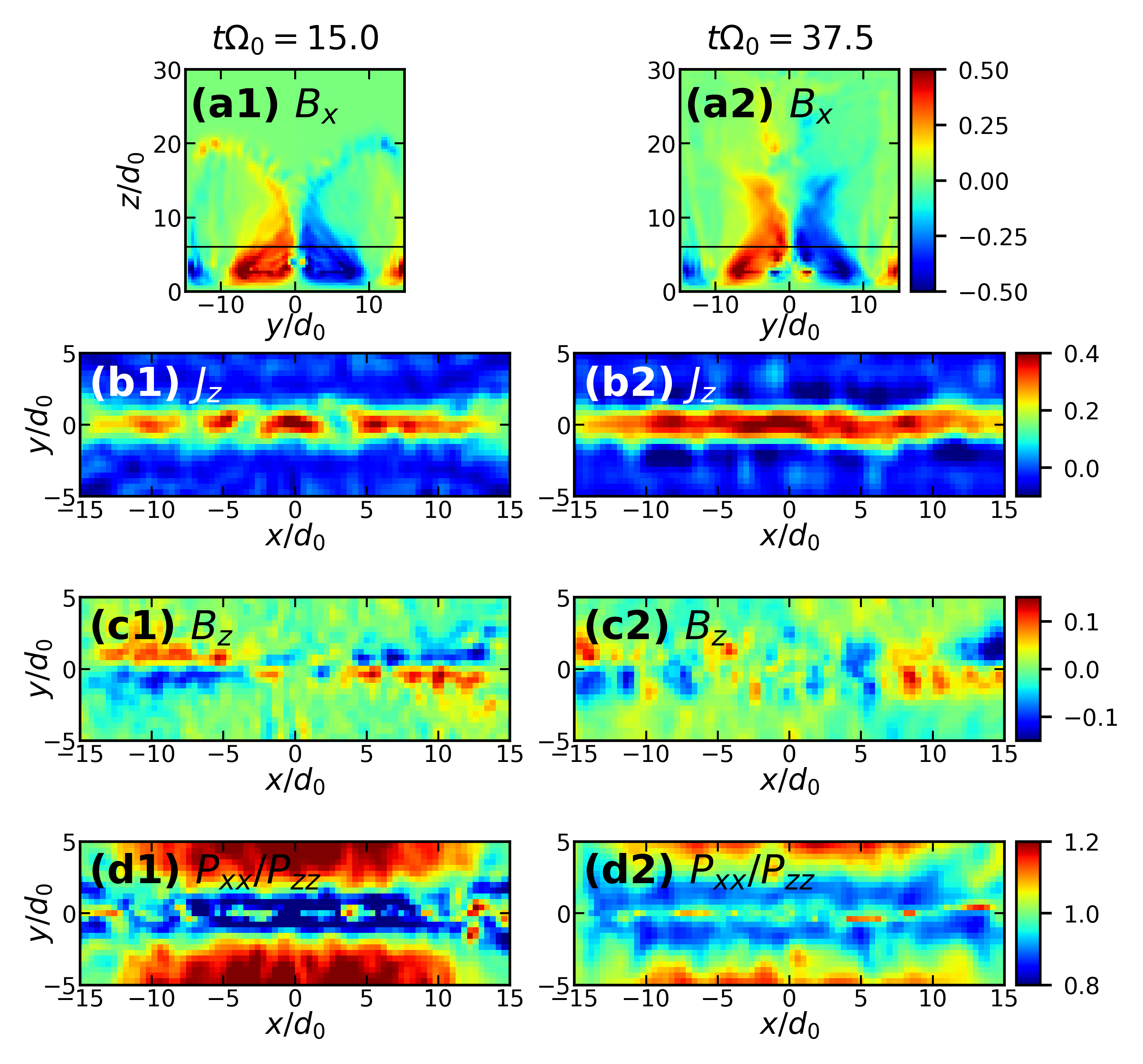}
\caption{Hybrid modeling of the reconnection dynamics for $S \sim 100$: comparison between the early ($t\Omega_0=15$) and late ($t\Omega_0=37.5$) stages. Panels (a1–a2) show the reconnecting in-plane magnetic field $B_x$. Panels (b1–b2) present the out-of-plane current density $J_z$. Panels (c1–c2) display the out-of-plane magnetic field $B_z$. Panels (d1–d4) show the ratio of the electron pressure tensor components $P_{xx}/P_{zz}$. Together these diagnostics demonstrate how the current sheet instability couples to the electron pressure anisotropy and modifies the magnetic topology in the nonlinear phase. Panel (a) corresponds to a central cut in the $y$–$z$ plane, while panels (b)-(d) show horizontal cuts at $z = 6 d_0$ (indicated by the black line in panels a1–a2).}
\label{fig:model}
\end{figure}

Figure~\ref{fig:model} shows the comparison between the plasma and fields parameters at two moments, i.e. during the   initial phase where the plasmas collide (corresponding to 1-2 ns of the experiment) and at late times (3.5-4.5 ns of the experiment). The field diagnostics  provide additional insight into the underlying dynamics. As shown in Figure~\ref{fig:model}.(a1-a2), the reconnecting magnetic field $B_x$   results, due to the strong compression induced by the in-flows, in a narrow out-of-plane current $J_z$ (shown in Figure~\ref{fig:model}.(b1-b2)) which  develops modulations consistent with tearing. As shown in Figure~\ref{fig:model}.(c1-c2), the Hall magnetic field $B_z$ displays a quadrupole structure at early times, but this ordered pattern breaks down into chaotic fluctuations once nonlinear reconnection sets in. Finally, the electron pressure tensor ratio $P_{xx}/P_{zz}$ (shown in Figure~\ref{fig:model}.(d1-d2)) reaches values of $\sim 0.8$, demonstrating that strong anisotropy builds up during sheet formation and remains significant even after plasmoid saturation. Together, these features show that both pressure anisotropy and Hall effects are strong, contributing  to the destabilization of the sheet and  shaping  its nonlinear evolution.

\begin{figure}
\includegraphics[width=0.48\textwidth]{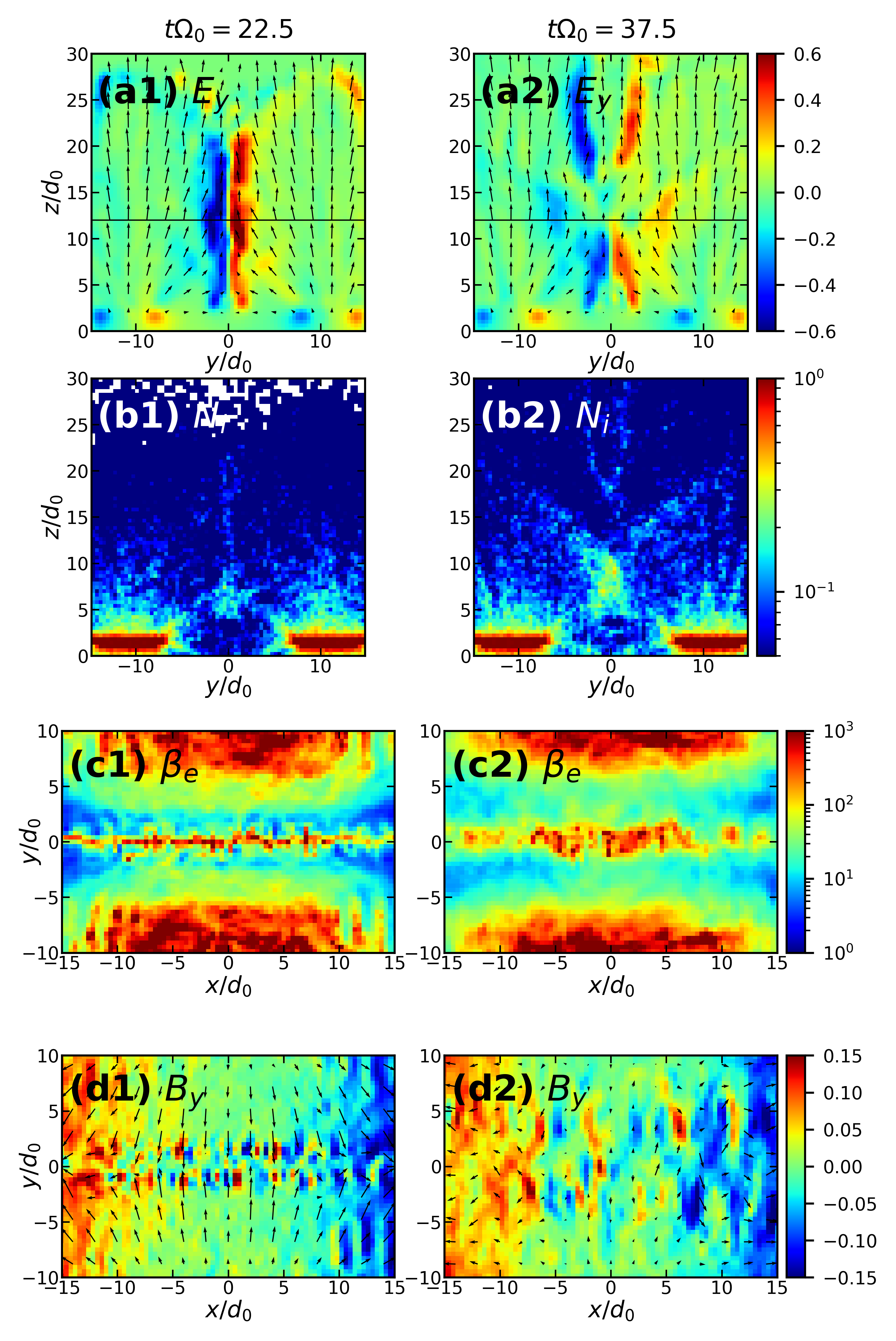}
\caption{Hybrid modeling of the reconnection dynamics for $S \sim 100$: comparison between middle ($t\Omega_0=22.5$) and late ($t\Omega_0=37.5$) stages. Panels (a1–a2) present the inflow electric field $E_y$ with superposed ion flow velocities in black. Panels (b1–b2) display the ion density $N_i$. Panels (c1–c4) show the ratio of electron pressure to the magnetic pressure, plasma $\beta$ parameters. Panels (d1–d2) present the in-plane magnetic field $B_y$ with superposed ion flow velocities in black. Panels (a) and (b) correspond to central cuts in the $y$–$z$ plane, while panels (c) and (d) show horizontal cuts at $z = 12 d_0$ (indicated by the black line in panels a1–a2). }
\label{fig:model2}
\end{figure}

Figure~\ref{fig:model2} compares the plasma and field structures during the middle stage of reconnection (corresponding to $\sim$2–3 ns in the experiment) and at a later stage ($\sim$3.5–4.5 ns). At intermediate times, the inflow ion velocity (black arrows in panels (a1–a2)) is strongly modified by the electric field $E_y$, which is itself generated by the enhanced electron pressure gradients (c1–c2), as the plasma is being compressed~\cite{sladkov2021}. The evolution of the density distribution (b1–b2) clearly shows plasma accumulation in the mid-plane, followed by redirection away from the neutral plane as reconnection proceeds. The consequences of these dynamics are twofold. First, the modification of the ion flows feeds back on the electric field and pressure balance, further amplifying localized instabilities. Second, the magnetic field $B_y$ (see panels (d1–d2)) reveals the transport of small-scale magnetic modulations away from the mid-plane, carried by the reflected ion flows. This transport reduces the magnetic compression at higher $z$, leading  at late times to the thinner central pattern observed in both the experimental and  synthetic proton radiographs, compared to the broader, mouth-like features observed at early and mid-times. %Also, the magnetic field $B_y$ is responsible for the redirection of the probing protons, clearly shows how small scale modulations are transported further from the neutral plane by reflected ion flows. 
Altogether, Figure~\ref{fig:model2} illustrates how plasmoid formation mediates the transition from a compressed current sheet to a fragmented, filamentary structure characteristic of the late nonlinear stage of reconnection.

\begin{figure}
\includegraphics[width=0.48\textwidth]{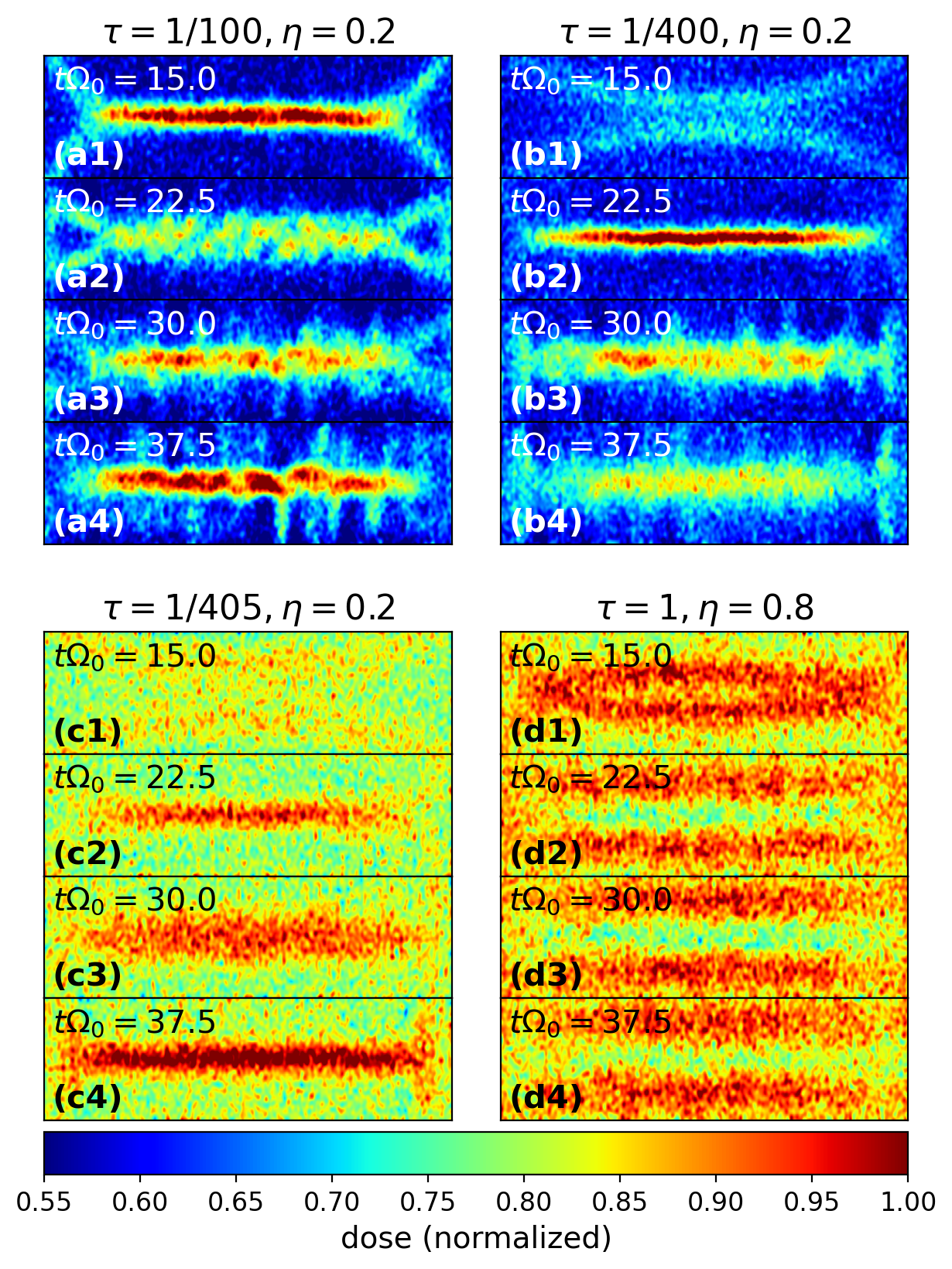}
\caption{Comparison of synthetic proton radiography images at four successive time steps for models with different dissipation parameters. Panels (a1–a4) correspond to $\tau = 1/100$, $\eta = 0.2$; (b1–b4) to $\tau = 1/400$, $\eta = 0.2$; (c1–c4) to $\tau = 1/405$, $\eta = 0.2$; and (d1–d4) to $\tau = 1$, $\eta = 0.8$. The parameter $\tau$ denotes the characteristic isotropization time of the electron pressure tensor~\cite{sladkov2021}, controlling the relaxation of pressure anisotropy, while $\eta$ represents the (uniform) resistivity level and thus the strength of collisional magnetic diffusion. Decreasing $\tau$ enhances anisotropy relaxation and progressively modifies the development of tearing structures, whereas increasing $\eta$ broadens the current sheet and suppresses plasmoid formation, leading to smoother proton fluence patterns.}\label{fig:proto_damp}
\end{figure}

Figure~\ref{fig:proto_damp} explores how the dissipation mechanisms alter the proton radiography patterns. When isotropization is artificially increased to $\tau=1/100$, the instability is slowed but not suppressed: the resulting images still display plasmoid chains very similar to the reference case (compare Fig.~\ref{fig:proto_damp}(a1-a4) to Fig.~\ref{fig:proto_exp}(b1-b4)). For stronger isotropization ($\tau=1/400$ and $1/405$, see Fig.~\ref{fig:proto_damp}(b1-c4)), the plasmoid structures become progressively less regular and intermittent. By contrast, the resistive case ($\tau=1$, $\eta=0.8$) shown in  Fig.~\ref{fig:proto_damp}(d1-d4) produces a diffuse sheet and smooth proton patterns, very unlike the experimental observations. This indicates that resistivity acts to broaden the current layer and suppress tearing. These comparisons crucially confirm that while resistivity strongly damps the instability, moderate isotropization—even stronger than in the main run—still permits robust plasmoid formation.

An important limitation of our work arises from the finite spatial and temporal resolution available in both the experimental diagnostics and the hybrid modeling. In the experimental case, the  magnetic field deflections inducing the proton radiographs are integrated along the probe path, which smooths, over sub-ion scales, structures that may be critical to observe  the onset of tearing and plasmoid growth. Similarly, the hybrid simulations, while reproducing the large-scale evolution of the current sheet, do not fully resolve the electron spatial scales. As demonstrated in recent comparative studies of the thermal Weibel instability using full and hybrid Particle-in-Cell (PIC) simulation approaches~\cite{sladkov2023}, under-resolving the electron physics can lead to artificially enhanced growth rates and accelerated coalescence of magnetic structures. This suggests that, while the presence of plasmoids in our results is robust, their quantitative properties—including growth rate, number, and detailed dynamics may differ from those in a fully resolved kinetic system. Future work should therefore focus on improving the resolution of both diagnostics and simulations, or on developing more advanced electron closure models, in order to fully capture the multi-scale nature of plasmoid-mediated reconnection.

\section{Summary and conclusion} \label{sec:conc}

The evidence we have here presented provides  new insights into magnetic reconnection, by exploring a previously uninvestigated regime characterized by driving two plasmas and their associated magnetic fields using long-pulse ($5$~ns), moderate-intensity ($10^{14}~\text{W/cm}^2$) lasers irradiation copper targets in a highly anisotropic geometry. In contrast to Pearcy et al.~\cite{pearcy2024}, who observed plasmoids in short-pulse, high-$\beta$ hemispherical bubble collisions, our setup allows us to probe the formation and fragmentation of the current sheet under conditions of extended drive and high aspect ratio. We are thus able to study the onset of reconnection, the  stability of the current sheet, and plasmoid evolution in a regime more relevant to structured astrophysical outflows and long-duration laboratory systems.  Moreover, we have established  a direct link between experimental observations, hybrid simulations, and theoretical predictions, providing a comprehensive picture of reconnection in high-energy-density plasmas. 

Overall, the good correspondance between the experiment and the hybrid modeling supports the idea that the interplay between electron pressure anisotropy and reconnection-driven ion flows is central to the destabilization and nonlinear evolution of the current sheet. The temporal evolution captured in Figures~\ref{fig:model} and~\ref{fig:model2} shows the transition from a thin, laminar sheet to a fragmented, filamentary structure consistent with plasmoid formation. Pressure anisotropy remains significant throughout the reconnection process, highlighting its role as a persistent driver of tearing instabilities. At the same time, the integrated nature of the proton radiography diagnostic and the finite resolution of the simulations imply that quantitative details—such as the number, size, and precise dynamics of individual plasmoids—cannot be directly inferred from the simulations, since plasmoids may coalesce or break up at different heights along the current sheet. These results also confirm that hybrid modeling provides a robust framework for interpreting the large-scale features of plasmoid-mediated reconnection observed in the experiment, while pointing to the need for higher-resolution studies to fully resolve the electron-scale dynamics and the multi-scale interactions. 

%Altogether, our results complement and extend recent work by providing a new experimental platform 

\section{Appendix A. Numerical model} \label{sec:model}

We use the hybrid code AKA~\cite{sladkov2020, akaGitHub}, build on classical principles~\cite{winske2003} of previous codes like
HECKLE~\cite{smets2011}, which keeps the ion description at the particle level and consider electrons as a fluid described by the pressure tensor~\cite{sladkov2021}. In this work, electrons are described in a fluid way, by the ten-moment model. Those ten moments are density - $n$ (equal to the total ion density by quasi-neutrality), bulk velocity - $\mathbf V_e$ and the six-component pressure tensor - $\mathbf P_e$. The electromagnetic fields are treated in the low-frequency (Darwin) approximation, as we consider that the phase velocity of electromagnetic fluctuations is small compared to the speed of light. Neglecting the displacement current, we then write an electron Ohm's law~:

\begin{equation}
\mathbf E = -\mathbf V_i \times \mathbf B + \frac{1}{e n}(\mathbf J \times \mathbf B - \boldsymbol{\nabla} . \, \mathbf P_e) + \eta \mathbf J \label{ohm}
\end{equation}

In Eq.~(\ref{ohm}), $\mathbf V_i$ is the ion bulk velocity, $n$ is the electron density, $\mathbf J$ is the total current density equal to the curl of $\mathbf B$. We use the explicit subcycling integration scheme for the six-component pressure tensor evolution equation~\cite{sladkov2021}. Electromagnetic fields are calculated on two staggered grids using a predictor-corrector scheme~\cite{winske1986}. The dynamics of the ions is solved using a first-order interpolation of the electromagnetic field~\cite{boris1972}. Resistive effects are included via a  resistivity parameter $\eta$ = 0.2, which corresponds to a Lundquist number $\sim 100$, that is constant in time and uniform.

In the numerical model, the magnetic field and the density are normalized to $B_0$ and $n_0$ respectively, times are normalized to the inverse of ion gyrofrequency $\Omega_0^{-1}$ (calculated using the magnetic field $B_0$), lengths are normalized to the ion inertial length $d_0$ (calculated using the density $n_0$), and velocities are normalized to the Alfv\'en velocity $V_0$ (calculated using $B_0$ and $n_0$). Mass and charge are normalized to the ion ones. The normalization of the other quantities follows from these ones. %Resistivity is a uniform constant in time.

The simulation domain is a 3D rectangular box with sizes $L_X = 40$, $L_Y = 30$, $L_Z = 30$. We use a 100$\times$75$\times$75 grid corresponding to a mesh size equal to 0.4$d_0$ in all directions, the time-step is $5\times10^{-3}$. We model the copper plasmas as having an average mass of 64 and charge of 19. We use free boundary conditions in all directions, a damping layer for the evolution equations (Faraday's law and pressure tensor evolution equation) and an outflow boundary for particles. For the cyclotron term integration in the pressure tensor evolution equation we use a ion-to-electron mass ratio $\mu$ = 100; neglecting the heat flux, we use an isotropization operator~\cite{sladkov2021} with characteristic time one ion scale $\tau$ = 1.  Consistently with the experiments, we set a pair of laser-produced plumes on top of a flat target, as shown in Fig.~\ref{fig:init}. The ablation operator produces linear heating of the dense target, with the electron temperature set equal to the ion one, $T_e=T_i=1$. The focal spots are elliptic with long axis 20$d_0$ and short 4$d_0$, resulting in a curvature radius at the contact point of 100$d_0$. The total simulation time is 37.5$\Omega_0^{-1}$, which corresponds to 5 ns of experiment for $B_0$ = 200. The self-generated magnetic field near the target surface stays on the level of 0.5, which is in accordance with the estimated experimental magnetic field in the focal spot $\sim$ 100 T.

\section{Appendix B. Plasmoids instability analysis} \label{sec:app}

\begin{figure}
\includegraphics[width=0.45\textwidth]{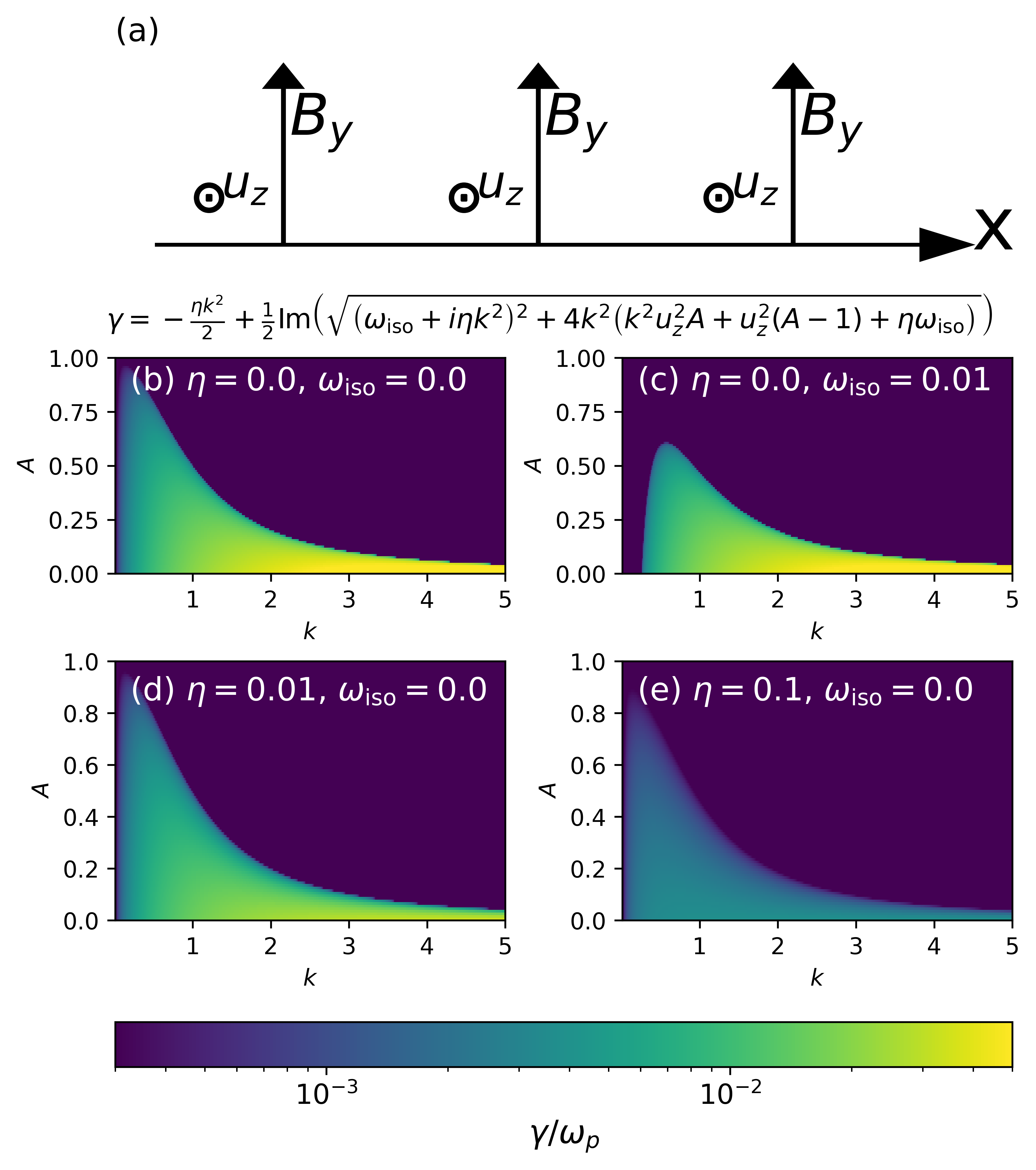}
\caption{Linear instability analysis of plasmoid formation in anisotropic plasmas. (a) Schematic of the configuration, where we have an initial magnetic field $B_y$ and an electron flow velocity $v_z$. (b–e) Normalized growth rate $\gamma/\omega_p$ as a function of the wavenumber $k$ and of the pressure anisotropy parameter $A = P_{xx}/P_{zz}$ for different normalized resistivities $\tilde{\eta}$ and isotropization frequencies $\tilde{\omega}_{\text{iso}}$, for a fixed thermal velocity $u_z = 0.02c$. The plots show that a moderate anisotropy favors the instability, while increasing either the resistivity or the isotropization damps the instability growth. These results highlight the competing roles of anisotropy-driven tearing and dissipative stabilization mechanisms.}\label{fig:appendix}
\end{figure}

We consider a 1D case where we have a constant uniform resistivity  $\eta$ and an electron pressure anisotropy which is balanced by the isotropization operator. The isotropization mocks an isotropization effect of the neglected divergence of the electron heat flux in the pressure tensor evolution equation. The initial magnetic field $B$ is in the $y$ direction (it is the one that, being modulated, results in the observations, through proton-radiography, of the plasmoids) and the electron flow velocity $v$ is along $z$, i.e. along the reconnection electric field $E_z$ (see Fig.~\ref{fig:appendix}.(a)). The ions are immobile,  for simplicity.
\begin{align}
\partial_x B_y &= -\mu_0 e n v_z \label{1} \\
\partial_t B_y &= \partial_x E_z \label{2} \\
E_z &= - \frac{1}{e n}\partial_x P_{xz} + \frac{\eta}{\mu_0} \partial_x B_y \label{3} \\
\partial_t P_{xz} &= -P_{xx} \partial_x v_z - \frac{e}{m_e} B_y (P_{xx} - P_{zz})  - \omega_{\text{iso}} P_{xz} \label{4}
\end{align}
In the generalized Ohm's law for the single component $E_z$,  we keep part of the divergence of the electron pressure tensor component $P_{xz}$ and the resistivity effect. In the evolution equation for the electron pressure tensor, we keep one term of the driver part proportional to the velocity gradient and cyclotron term as proportional to the electron gyrofrequency, while the damping due to the isotropization operator is proportional to the isotropization frequency  $\omega_{\text{iso}}$, as is often used for the purpose of numerical stability in modelling~\cite{sladkov2021}. From a physical point of view, the operator can represent the isotropization due to the electron-electron collisions effect. We assume that all perturbed quantities vary as $\sim e^{i k x - i \omega t}$. The linearized system of equations becomes:

\begin{eqnarray}
i k B_y &=& -\mu_0 e n v_z  \quad \Rightarrow \quad v_z = -\frac{i k}{\mu_0 e n} B_y  \label{eq:linear_bfield} \\
-i \omega B_y &=& i k E_z \quad \Rightarrow \quad E_z = -\frac{\omega}{k} B_y \label{eq:faraday} \\
E_z &=&   -i \frac{ k}{e n} P_{xz} + \frac{\eta}{\mu_0} i k B_y \label{eq:ohm}
\end{eqnarray}
\begin{align}
-i \omega P_{xz} &=& -P_{xx} i k v_z  
 - \frac{e}{m_e} B_y (P_{xx} - P_{zz}) - \omega_{\text{iso}} P_{xz} \label{eq:pxz_evol}
\end{align}

Plugging into Ohm’s law gives:

\begin{align}
 - \frac{\omega}{k} B_y = -i \frac{ k}{e n} \left( \frac{ i k P_{xx} \frac{i k}{\mu_0 e n} B_y  - \frac{e}{m_e} B_y (P_{xx} - P_{zz})}{-i \omega + \omega_{\text{iso}}} \right) + \frac{i k \eta}{\mu_0} B_y
\end{align}

We use the normalization:
\begin{align*}
\omega_p^{-2} &= \frac{m_e \epsilon_0}{e^2 n} = \frac{m_e}{\mu_0 c^2 e^2 n} ,\quad u_z^2 = \frac{P_{zz}}{n m_e c^2}, \quad A = \frac{P_{xx}}{P_{zz}} \\
\tilde{\omega} &= \frac{\omega}{\omega_p}, \quad \tilde{k} = \frac{c k}{\omega_p}, \quad \tilde{\eta} = \frac{\eta \omega_p}{\mu_0 c^2},\quad \tilde{\omega}_\text{iso} = \frac{\omega_\text{iso}}{\omega_p}
\end{align*}

\begin{equation}
\omega^2  +  ( \omega_{\text{iso}} + i \eta k^2 ) \omega -  (  A  u_z^2 k^2 +    ( A - 1)u_z^2
+ \eta\omega_{\text{iso}})k^2
= 0
\end{equation}

Then the solution is:
\begin{align}
\omega &= \frac{-\left( \omega_{\text{iso}} + i \eta k^2 \right)}{2} \nonumber \\
&\quad \pm \frac{1}{2} \sqrt{
\left( \omega_{\text{iso}} + i \eta k^2 \right)^2
+ 4k^2 \left[ k^2 u_z^2 A + u_z^2 (A - 1) + \eta \omega_{\text{iso}} \right]
}
\end{align}

Therefore, the growth rate \( \gamma =\text{Im}(\omega) \) is:
\begin{equation}
\resizebox{\columnwidth}{!}{$
\gamma = \frac{-\eta k^2}{2} + \frac{1}{2} \text{Im}\left(
\sqrt{ \left( \omega_{\text{iso}} + i \eta k^2 \right)^2 
+ 4k^2 \left( k^2 u_z^2 A + u_z^2 (A - 1) + \eta \omega_{\text{iso}} \right) } 
\right)
$}
\end{equation}

the zero resistivity and zero isotropization cases are in accordance with what was  previously found as solution for  collisionless plasmas~\cite{sladkov2023}.

Figure~\ref{fig:appendix}.(b-e) shows the normalized growth rate \(\gamma / \omega_p\) as a function of the wavenumber \(k\) and the pressure anisotropy parameter \(A = P_{xx} / P_{zz}\), for different values of the normalized resistivity \({\eta}\) and of the isotropization frequency \({\omega_{\text{iso}}}\), with fixed flow velocity \(u_z=0.02\). As can be seen, the instability is strongest at low values of  $A$ and for moderate \(k\), and it becomes progressively damped as \({\eta}\) or \({\omega_{\text{iso}}}\) increase.

Both collisional resistivity and isotropization act as stabilizing mechanisms for the tearing instability in anisotropic plasmas. The inclusion of finite resistivity or isotropization frequency in the linear model leads to damping of the instability growth rate, as shown analytically and confirmed in parameter scans (see Fig.\ref{fig:appendix}). Simulations further validate this behaviour: resistivity results in diffuse current layers and smooth radiographs, while isotropization suppresses plasmoid formation. These findings highlight the role of dissipation and kinetic relaxation processes in controlling the transition between laminar Biermann battery mediated~\cite{matteucci2018} and plasmoid-mediated reconnection.

\section{Appendix C. Experimental plasma parameters measurements %of current sheet parameters with 
using Thomson scattering} \label{sec:thomson}

%During the experiment described in ~\ref{sec:exp}, we used several diagnostics to characterise the plasmas and the current sheet. One of these diagnostics was t
Time-resolved Thomson scattering spectrometry was performed in order to characterize the density and temperature of the plasma %. In experimental plasma physics, Thomson scattering diagnostic is an efficient method to measure fundamental plasma parameters like the electron density n$_{e}$ and provide a direct insight of the temporal evolution of the electron temperature T$_{e}$ and the ion temperature T$_{i}$ 
~\cite{Froula1972}% during the laser-target interaction and the plasma heating
.
%It consists in an elastic scattering of a photon $\textit{hv}$ and a charged free particle (electron or proton) while staying in the classical electromagnetic description. In this regime, the photon is considered as an electromagnetic wave, so during the interaction with the particle, the electric and magnetic components of the the wave apply on it a Lorentz force that sets it in motion. Energy is transferred to the particle, that emits an electromagnetic radiation. We 
The diagnostic used an auxiliary laser (20 J, 5 ns duration, 526.5 nm wavelength) %use an external probe 
to irradiate the plasma  and scatter on the plasma waves, with a variable delay compared to the lasers irradiating the solid targets to drive the reconnection. % the particles to look at the light remitted.
%Thomson scattering spectrometry is a reliable diagnostic to collect these information as a laser delivering a few joules is enough to collect data ( E = 50 J in the case of the experiment). 
 The collection of the scattered light was set at 90° from the direction of the Thomson probe laser, such that we were % We set our diagnostic such as collecting information for both electrons and ions. The condition to achieve that is to be 
in a regime of collective scattering. %from electrons correlated with the motion of the plasma waves, and allowing the fitting of ion acoustic waves too. The spectral resolution of the probe must be adapted to reach the collective regime, as one needs to have $\alpha > 1$, with \alpha = 1/(\textbf{k}\lambda$_{D}$), where $\textbf{k}$ is the wave vector of the scattering probe and \lambda$_{D}$ is the Debye length ~\cite{Funaba2022}.

\begin{figure}
    \centering
    \includegraphics[width=1\linewidth]{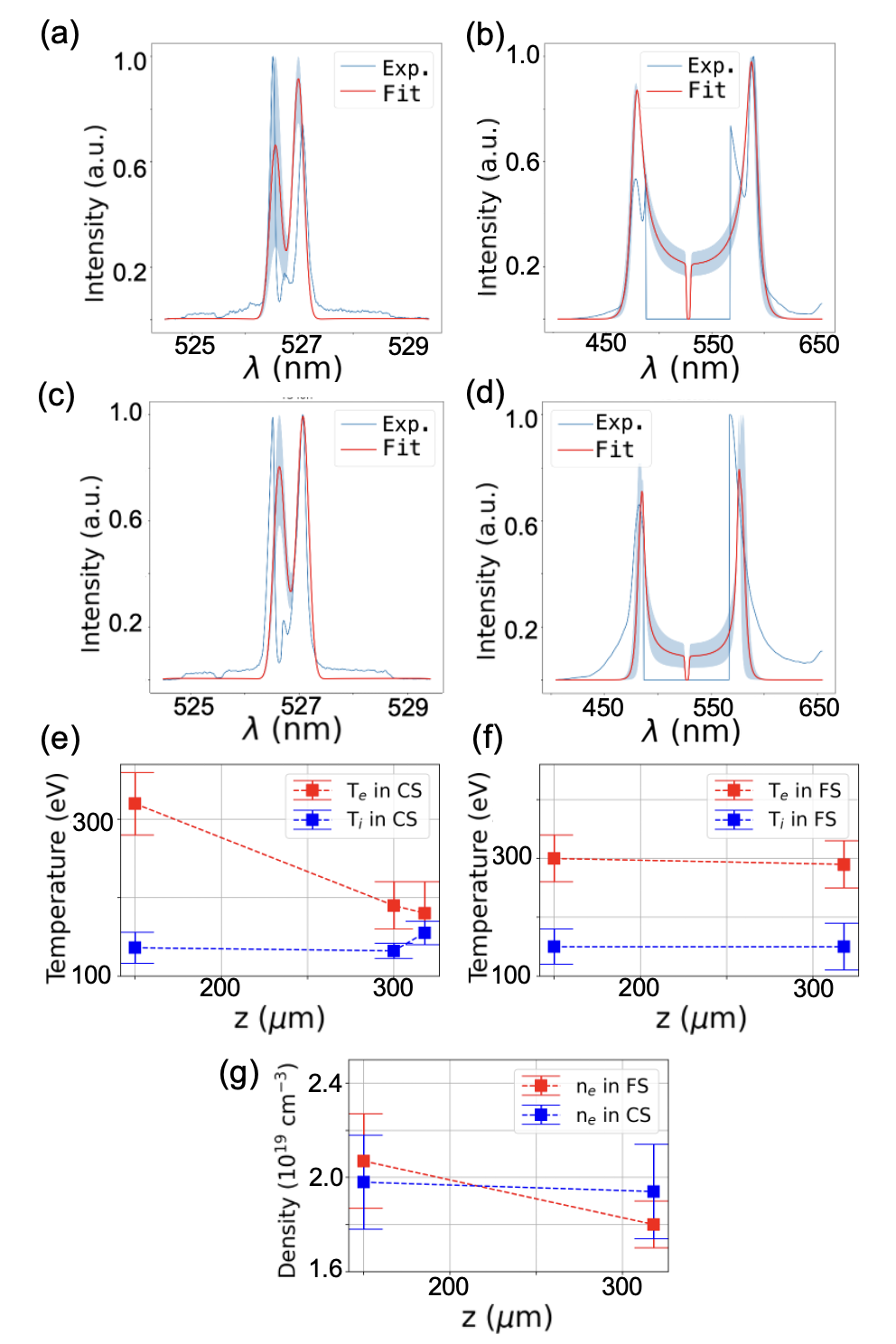}
    \caption{(a-d) Fitted Thomson scattering spectra (ion spectra on the left and electron spectra on the right) for two spatial positions alongside the z-axis (normal to the reconnection plane) within the current sheet : 150 $\mu$m away from the targets surface (a-b) and 318 $\mu$m away (c-d), respectively measured at 1.95 ns and 1.83 ns. The aspect of the spectra is dependent on variations of T$_{i}$ and T$_{e}$. The blue region around the theoretical fit represents the variations of the spectra in intervalles of uncertainties around mean values T$_{e}$ and T$_{i}$, represented by the red fit. (e) Profiles of T$_{e}$ and T$_{i}$ measured in the current sheet.(f) Profiles of T$_{e}$ and T$_{i}$ measured in the axis of the laser focal spot. (g) Profiles of n$_{e}$ measured in the current sheet ("CS") and in the axis of the laser focal spot ("FS").}
    \label{fig:thomson}
\end{figure}

The analysis of the scattered light spectra, shown in Fig.~\ref{fig:thomson} indicates that at the probed locations, %A preliminary analysis give us an experimental insight of an average 
the electron density is %of 
$n_{e} = 2 \times 10^{19}$ cm$^{-3}$. %Further analysis will reveal only small variations of this density during the different shots. 
Together with the temperatures that are deduced from fitting the spectra, as shown in Fig.~\ref{fig:thomson}, we can infer %With the Knowing that, we can establish our table of ionisation for the copper targets, and integrate these values to our numerical analysis of the spectrograms, finding 
an ionisation level of the copper of Z = 14-% or Z = 
15.%, depending on the shot 

%Fig.~\ref{fig:thomson}.(a) and (c) presents fits of Thomson scattering spectrograms showing narrow and well-defined ion-acoustic peaks. To extract the I($\lambda$) spectra, the spectrograms were integrated over a selected temporal window shorter than 0.5 ns. This duration is conscientiously shorter than the total time window (10 ns) to place the measure in quasi-stationary plasma conditions within a spatial interval around the probe wavelength $\lambda_{probe}$ = 527 nm. The intensity is normalized on each spectrum. 
Fits were performed on different spectra %ograms obtained during the experiment in order to evaluate the temporal evolution of the temperature in the current sheet as well as of the spatial evolution, as 
that were recorded by having the probe laser %was 
aligned at locations: we not only positioned  the probing volume (defined by the intersection of the diagnostic line-of-sight with the probing laser axis) either within the current sheet (i.e. at x=y=0) or above the laser focal location (i.e. at x=0 and y=250  $\mu$m), but we also probed the plasma at different height, along the z-axis, namely (100, 300 and 325 $\mu$m). %different distances from the main targets from shot to shot.
The analysis of the Thomson ion spectra (Fig.~\ref{fig:thomson}(a-c)) provides the main information about the electron and ion temperatures and levels of ionization, while %but a complete evaluation of the plasma parameters also requires 
the analysis of the Thomson electron spectra (Fig.~\ref{fig:thomson}(b-d)) allows to pinpoint the electron density in the plasma. %Consequently, the same fitting method was applied for electron Thomson scattering results, after removing the central narrow ion peaks of the spectrograms.

Overall, the results show that %The Thomson Scattering diagnostic gives the plasma thermodynamic state within the current sheet ~\cite{Finch2021}. A
as soon as the plasmas start to expand from the laser irradiation zones and interact in the reconnection zone, the electrons are heated considerably more than the ions, especially in the core of the current sheet. Indeed, due to the electric field in the reconnection sheet and as the electrons have a high capacity of conducting thermal energy, they are preferentially heated compared to the heavier ions with weaker thermal conduction. Electrons located in a high-$\beta$ regions are also subject to high thermal pressures and sharp temperature gradients, and as such  are heated in a time shorter than the electron-ion exchange time $\tau_{ei}$. 

Besides being correlated to the $\beta$ ratio, T$_{e}$ also fluctuated due to the internal activity of the current sheet and the transition to the plasmoid-dominated regime. During this regime shift, particles confined by the local magnetic field around plasmoids are heated more significantly than those in between due to the high locally increasing amplitude of the $\beta$ ratio inside of the plasmoids ~\cite{Numata2015}. %\textcolor{red}{Electrons accumulate locally in plasmoids when they get formed but get ejected quickly, establishing a continuous flux of electrons through the reconnection zone and ejected from the sheet. Indeed, the electron densities measured at the probed locations (Fig.~\ref{fig:thomson}(b-d)) remain stable, and correspond to the density in the focal spot axis at a distance z = 150 $\mu$m. However, the density does decay with the distance in this zone as it is expected for a laser-driven plasma, reaching n$_{e}$ = 1.7 \times 10$^{19}$ cm$^{-3}$.}

Unlike T$_{e}$, T$_{i}$ remains stable as the process takes place during the laser impulsion. During the plasmoids regime, the average value of $\beta$ and T$_{e}$ decreases with distance, even if this occurs while the laser pulse produces continously hot electrons (Fig.~\ref{fig:thomson}(f)). Overall, this %As a 
results in a %, the 
ratio of temperatures %is 
T$_{e}$/T$_{i}$ $\gg$ 1 (see Fig.~\ref{fig:thomson}(e)) in the current sheet and close to the targets. However, we can also observe that this ratio decreases %and strives 
toward 1 as we move away along the z-axis %with distance 
from the core of the reconnection region, showing a progressive electron-ion thermal equilibrium in lower $\beta$ zones.

\section{Appendix D. Optical pyrometry evidence for the fragmentation of the current sheet at late times} \label{sec:pyrometry}

%\textcolor{red}{Proton radiography is a diagnostic used to characterize the deflection of protons by magnetized plasmas. The figure of deflection is screened on a stack of several radiochromic films (RCF), a dose dependent detector ~\cite{Borghesi2004}. During the experiment, this method was used with a 1 picosecond-long TNSA proton beam that probed the reconnection structure. Each proton will deposit its energy on one specific film, as each layer corresponding to one discrete proton energy range. The magnetic structure is visible on the RCF placed in the trajectory of the TNSA proton beam as they darken when exposed to radiation ~\cite{Borghesi2003}. As the TNSA pulse is considerably shorter than the time-resolution of the magnetic field topology during the shot, the reconnection structure screened on the RCF appears frozen in time in the proton beam referential. With this technique it is possible to modify the delay between the long pulse lasers and the proton beam and picture from shot to shot the time evolution of the current sheet.}

\begin{figure}
    \centering
    \includegraphics[width=1\linewidth]{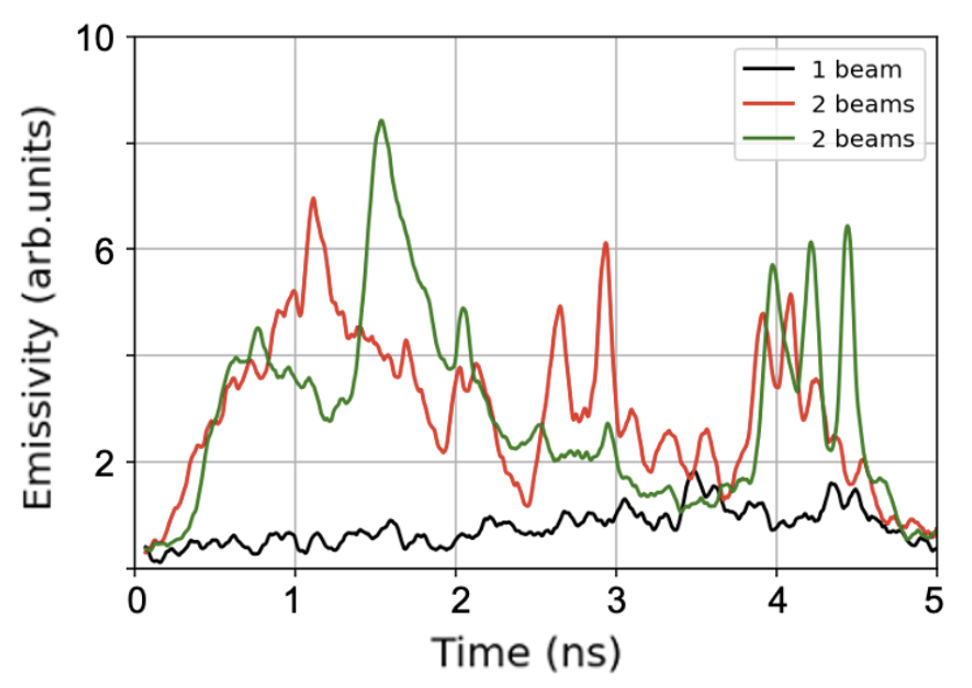}
    \caption{%(a) Experimental RCF of two distinct shots in same conditions at t = 4.3 ns showing the repeatability of the plasmoid occurrence in the elongated sheet configuration. (c) 
    Temporal lineouts of optical pyrometry measurements, integrated along the z-axis, of plasma self-emission  as a function of time, induced by the plasma heating in the reconnection region (at x=y=0) in the 2 beams configuration (reconnection occuring) and in the 1 beam configuration (no reconnection). The starting time t = 0 ns corresponds to the start of the laser irradiation of the targets.}
    \label{fig:RCF}
\end{figure}

%\textcolor{red}{Fig.~\ref{fig:RCF}(a) shows the temporal evolution of the reconnection structure as captured by proton radiography for different proton energy ranges. This diagnostic permits to track experimentally the different phases of the current sheet and the transition of regimes as described in ~\ref{sec:exp}. At early times, RCF reveal a thin and stable current sheet before it starts getting distraught and widen into a mouth shape around t = 2 ns. At t = 4.3 ns, the plasmoids are fully established. This transition between regimes induces a dissipation of magnetic energy from the sheet, evidenced by the protons deflected all around the current sheet starting at t = 3 ns.} 

%\textcolor{red}{The formation of plasmoids is recurring in an elongated sheet due to the fragmentation of the plasma and the growing instabilities in the sheet during the reconnection process. Fig.~\ref{fig:RCF}(b) shows experimental proton radiographies of two different shots performed in the same conditions. In both cases, reconnection occurs between the plasmas magnetized only via the Biermann battery effect and plasmoids get formed at later times of the reconnection. Plasmoid formation is consequently a repeating mechanism of magnetic reconnection in the elongated sheet configuration and is possible to observe in the laboratory in presence of only self-generated magnetic fields.}

The temporal evolution of the emissivity, in the optical visible domain,  of the plasma observed at the location of  the current sheet (at x=y=0) is measured  using an optical pyrometry imaging setup, coupled to a streak camera~\cite{Kuramitsu2018}. %The optical diagnostic is used here to address the proton radiography results. 
 The time-resolved plasma emissivity was integrated along the line of sight, which coincided with the z-axis.

We observe, as shown in Fig.~\ref{fig:RCF}, that the emissivity behaves strongly differently when there is  either only one plasma, or in presence of the two colliding plasmas. As discussed in our previous work\cite{Bolanos2022}, the emitting plasma is in an optically thin regime, meaning that the self-emission increases with the plasma density, but decreases with the plasma temperature. Consistently with what we previously observed \cite{Bolanos2022}, when only one plasma is present, the plasma emissivity remains low and quasi-constant during the whole plasma lifetime. This just reflects the lateral expansion of the hot plasma from the laser spot onto the x=y=0 location (250 $\mu$m away from the laser spot), and the constant density and temperature of the laser-driven plasma at that location in the absence of a reconnection process. However, the plasma emissivity display a very different behaviour in the presence of two reconnecting plasmas (two shots are shown for reproducibility).  In our previous experiment using a X-line configuration of two hemispherical reconnecting plasmas \cite{Bolanos2022}, we observed that the plasma emissivity was also strongly increasing initially (compared to the single plasma case), due to the collision between the two plasmas inducing an increased density in the reconnection region. This is also what is initially observed here. However, the later behavior differs from what we observed when we used a X-line configuration \cite{Bolanos2022}. Indeed, in that case, we observed that the plasma emissivity was steadily decreasing due to the increased temperature of the plasma in the current sheet as the magnetic energy is transferred to the plasma. Here, in our elongated configuration where the current sheet is fragmented in plasmoids, we observe, on top of the progressive slow decrease of the emissivity, strong jumps and oscillations of the emissivity. These jumps appear more strongly after 4 ns, coincidentally with the observation of current sheet fragmentation in the proton radiographs. %the first peaks of emissivity for the 2 beams configuration correspond to the stable regime when the current sheet remains thin and confined. This high emissivity is linked to the earliest stages of the reconnection, characterized by high values of electron temperature T$_{e}$ and density n$_{e}$, as well as strong parallel electric field E$_{//}$. The amplitude of the emissivity of the plasma decreases around 2 ns, corresponding to the time when the RCF show the beginning of the regime transition. Secondary peaks can be observed at this stage as the broadening of the sheet goes with energy releases. After 4 ns, final peaks are observed, during the plasmoid regime. 
However these peaks remain weaker than the first ones, as the plasma is then more extended and less dense, due to plasmoid formation and ejection~\cite{Zhao2022}. %Therefore, proton radiography and optical pyrometry results are correlated. It is noticed that 
This further supports the analysis of the plasma within the current sheet, since the temporal evolution of the emissivity of the plasma is governed and dominated by the reconnection topology derived from the proton radiography data. 

%\textcolor{red}{Furthermore, proton radiography reveals the emergence of a secondary reconnection process, orthogonal to the primary one~\cite{Liu2024}. This second current sheet does not occur between two distinct interacting plasmas but results directly from the fragmentation of the plasmas. The magnetic field carried by the plasmas undergoes perturbations from the main reconnection process and starts to reconnect with itself. This phenomenon can be observed at early times in the process and can develop alongside the rest of the structure, especially during the transition between the reconnection regimes. As the regime change induces a wider and more diffusive current sheet, the diffusion of the thermal energy is more isotropic. This heats colder regions outside of the current sheet, which will nurture the growth of secondary processes~\cite{dong2012}. The second reconnection sheet widens simultaneously with the main one and evolves toward the plasmoid regime. However, this orthogonal reconnection remains more diffused and less stable than the principal. The current sheet is less energetic and magnetic topologies obtained with RCF indicate that they lasts for short timescales, making it a complementary process of the first one.}

\bigskip

\section{Acknowledgments}\label{sec:acknowlegments}

We thank the crew of the LULI2000 facility for their excellent technical support. T.W. acknowledges the financial support of the IdEx University of Bordeaux / Grand Research Program "GPR LIGHT". This work was supported by the European Research Council (ERC) under the European Union’s Horizon 2020 research and innovation program (Grant Agreement No. 787539; J.F.), by the National Sciences and Engineering Research Council of Canada (NSERC) (Grant — RGPIN-2023-05459, ALLRP 556340 – 20; T.W., P.A.), Compute Canada (Job: pve-323- ac).

\section{Author Declarations - Conflict of interest}
The authors have no conflicts to disclose.

\section{Author Contributions}
J.F. and A.A. conceived the project. J.F., T.W., H.A., V.A. and W.Y. performed the experiment. J.F., T.W., A.A., I.C., W.Y. analyzed the data. A.S. performed the simulations. P.A. and E.d.H. provided guidance and funding. A.S., J.F., S.N.C. and T.W. wrote the paper. All authors commented and revised the paper.

\section{Data Availability}
The data that support the findings of this study are available from the corresponding author upon reasonable request.

\bibliographystyle{apsrev4-2}
\bibliography{biblio.bib}

\end{document}